\documentclass[aps,prb,twocolumn,showpacs,groupedaddress]{revtex4}
\pdfoutput=1
\usepackage{graphicx}
\usepackage{amssymb}
\usepackage{amsmath} 
\usepackage{amsbsy} 
\usepackage[latin1]{inputenc} 
\usepackage{mathrsfs}
\usepackage{ulem}
\usepackage{color}

\definecolor{red}{rgb}{1,0,0}
 
\begin{document} 
 
\widetext 
 
\title{Intrinsic exchange-correlation magnetic fields in exact current-density functional theory for degenerate systems} 
\author{J.~D.~Ramsden} \affiliation{Department of Physics, University of York, Heslington, York YO10 5DD, United Kingdom} \affiliation{European Theoretical Spectroscopy Facility (ETSF)}
\author{R.~W.~Godby} \affiliation{Department of Physics, University of York, Heslington, York YO10 5DD, United Kingdom} \affiliation{European Theoretical Spectroscopy Facility (ETSF)}
\date{\today}

\begin{abstract} 
We calculate the exact Kohn-Sham (KS) scalar and vector potentials that reproduce, within current-density functional theory, the steady-state density and current 
density corresponding to an electron quasiparticle added to the ground state of a model quantum wire.  Our results show that, even in the absence of an external 
magnetic field, a KS description of a steady-state system in general requires a non-zero exchange-correlation magnetic field that is purely mechanical in origin. 
The KS paramagnetic current density is not, in general, that of the interacting system in any gauge.
\end{abstract}
 
\pacs{71.15.Mb, 73.63.-b, 73.23.-b, 85.35.Be} 
\maketitle

\section{Introduction}

Time-dependent density-functional theory \cite{RungeGross1984} (TDDFT) in the Kohn-Sham (KS) scheme \cite{vanLeeuwen99} is a powerful and in principle exact tool for 
predicting the dynamics of nonequilibrium systems subject to time-dependent scalar potentials. The success of the theory lies in the casting of properties of 
interacting $N$-particle systems in terms of systems of $N$ noninteracting (Kohn-Sham) electrons, and the unique determination of KS potentials by the time-dependent 
charge density and the initial state. In the ground-state limit, the theory approaches the standard density-functional theory (DFT) of Hohenberg, Kohn and Sham 
\cite{HohenbergKohn1964,KohnSham1965}.

In the presence of time-dependent {\it vector} potentials, one needs to know how the potentials couple to the time-dependent current density. The current density can be 
decomposed into two parts: the longitudinal part obeying $\nabla\times\mathbf{j}_{\textrm{L}}(\mathbf{r},t)=\mathbf{0}$, and the transverse part obeying 
$\nabla\cdot\mathbf{j}_{\textrm{T}}(\mathbf{r},t)=0$. The longitudinal part is given entirely by the time-dependent density via the continuity equation
\begin{equation}
 \frac{\partial}{\partial t}n(\mathbf{r},t) + \nabla \cdot \mathbf{j}_{\textrm{L}}(\mathbf{r},t) = 0.
\end{equation}
The transverse part of the current is not given by the continuity equation and must be calculated directly from the time-dependent wavefunction. The complete current 
density is given by
\begin{equation}
 \mathbf{j}(\mathbf{r},t) = \mathbf{j}_{\textrm{p}}(\mathbf{r},t) + \mathbf{A}(\mathbf{r},t)n(\mathbf{r},t),
\end{equation}
where
\begin{equation}
 \mathbf{j}_{\textrm{p}}(\mathbf{r},t) = \left< \Psi(t) \left| \hat{\mathbf{j}}_{\textrm{p}}(\mathbf{r}) \right| \Psi(t) \right>
\end{equation}
is the paramagnetic current density operator, $\mathbf{A}(\mathbf{r},t)$ is the time-dependent external vector potential, $n(\mathbf{r},t)$ is the time-dependent 
density and $\left| \Psi(t) \right>$ is the time-dependent many-body wavefunction.

It has been shown \cite{dAgostaVignale05} that, even in the absence of external vector potentials in the time-dependent regime, the corresponding KS system will 
generally have an exchange-correlation (XC) vector potential that couples to the full current density. As such, one needs a time-dependent current-density functional 
theory (TDCDFT) to fully describe the dynamics of electronic systems. Such a theory exists \cite{GhoshDhara88} and is representable in a KS scheme \cite{Vignale04} 
in which the KS potentials are unique functionals of the initial state and the time-dependent physical current density.

For ground-state systems, two CDFTs based on the physical current have been proposed \cite{PanSahni2010,Diener1991} but neither fulfill the requirements of uniqueness 
and amenability to a KS minimisation scheme \cite{TellgrenEtal2012,Vignale13}. The most complete theory of ground-state current-carrying systems subject to external 
scalar and vector potentials remains the current- and spin-density functional theory (CSDFT) of Vignale and Rasolt \cite{VignaleRasolt1988} which takes as its basic 
variables the ground-state charge and paramagnetic current densities $(n, \mathbf{j}_{\textrm{p}})$. It has been shown that the ground-state wavefunction, and therefore 
the universal functional 
\begin{equation}
\label{eq:unifunc}
 F \left[ n, \mathbf{j}_{\textrm{p}} \right] = \left< \Psi \left| \hat{T} + \hat{U} \right| \Psi \right>,
\end{equation}
is uniquely defined by these quantities. However, at present there exists no current-density functional theory (CDFT) for ground-state systems to which TDCDFT approaches 
in the steady, ground-state limit. 

In the KS scheme for CSDFT, one then constructs a system of noninteracting electrons having the same charge and paramagnetic current density as the interacting system it 
represents. There are two problematic aspects of such an approach. First, the KS system will typically not have the same physical current as the system it represents 
and, as such, is not approached by TDCDFT in the ground-state limit. Different schemes for the construction of KS systems is the primary focus of this work.

Second, and also very pertinent to this study, the KS potentials are not determined by the two basic densities (in contrast to other KS-DFTs). One area of research in 
DFT that is quickly growing in activity is the calculation of exact KS systems from exactly-solvable systems for the purpose of advising the construction of better 
functionals, including the exact KS potentials required to reproduce the analytic two-particle wavefunction solutions of the time-dependent Schr\"odinger equation 
\cite{Thiele2008,Elliott2012}, the exact potentials for time-dependent Hubbard chains \cite{Verdozzi2008}, and the time-dependent potentials required to describe 
nonequilibrium quasiparticles described by a nonlocal model self-energy operator \cite{RamsdenGodby2012}. 

Here, we take a related approach to the steady-state regime. We consider a three-dimensional steady-state infinite wire whose QPs are, once again, taken to be described 
by the model self-energy operator employed  in Ref. \cite{RamsdenGodby2012}. The systems with which we concern ourselves are those in which the $N$ ground-state 
electrons fill the valence band together with the  standing state at the bottom of the conduction band, yielding a convenient nondegenerate ground state of zero current 
density. The $N+1^{th}$ electron is a QP added to one of the next-lowest-energy, degenerate, current-carrying Bloch states. As such, the system approximates a quantum 
wire bridging two electron reservoirs held at slightly different chemical potentials.

Degenerate systems (for which the DFT and VR existence proofs are not intended to hold) may have a current even in the absence of an external vector potential, making 
an interesting study. The external scalar potential of a system subject to no external vector potential is uniquely determined by $n(\mathbf{r})$ alone (or, more 
accurately, the density ensemble of all degenerate states \cite{Levy1982}) and thus is amenable to description by DFT. However, the ground-state wavefunction is uniquely 
defined by $(n, \mathbf{j}_{\textrm{p}})$ together, as in VR theory, and thus the system is also amenable to study within current-density functional theory (CDFT) for 
spinless systems. A key question is if and how the exact KS schemes differ between DFT, the CSDFT of Vignale and Rasolt, and a CDFT wherein the KS and interacting 
systems share the same physical current.

A second question, central to the study in this paper, concerns the construction of KS systems. Having chosen the basic variables that characterize a many-electron 
system, it does not necessarily follow that the KS representation will have the same values of other variables as the interacting system, which is particularly 
important if those other variables are measurable properties of the system. In particular, in order to have a bearing on the TDCDFT, 
it is necessary that the physical current density be reproduced by the KS system at all times. We investigate the implications of different rules for constructing 
current-carrying KS systems.

\section{The quasiparticle current density}

The wave equation governing the added QP is (in atomic units as throughout) 
\begin{align}
\label{eq:qpequation}
 \left( - \tfrac{1}{2}\nabla^2 + v_{\mathrm{ext}}(\mathbf{r}) + v_{\mathrm{H}}(\mathbf{r}) - E_{\mathrm{QP}} \right) \psi_{\mathrm{QP}}(\mathbf{r}) \nonumber \\ 
 + \int d^3\mathbf{r}'~\Sigma(\mathbf{r},\mathbf{r}') \psi_{\mathrm{QP}}(\mathbf{r}') = 0 
\end{align} 
where $E_{\mathrm{QP}}$ is the quasiparticle energy, $v_{\mathrm{ext}}(\mathbf{r})$ the external potential to which the entire system is subject, 
$v_{\mathrm{H}}(\mathbf{r})$ the Hartree potential, and $\Sigma$ the self-energy operator. Since the QP is to be added to the lowest-energy unoccupied state, the QP 
lifetime is expected to be infinite and therefore the self-energy real. Generally the self-energy is an energy-dependent operator, however it has been shown 
\cite{Godbyetal1988} that this energy-dependence yields rather small quantitative changes to the band structure in ground-state semiconductors and as such is not 
expected to give rise to any qualitatively different features in the corresponding KS potential. As such, we take the self-energy to be nonlocal but energy-independent 
and Hermitian.

The electrons are confined to the wire by the scalar potential 
\begin{equation} 
\label{eq:GSpot} 
 v_{\mathrm{ext}} + v_{\mathrm{H}} = Hr^6
\end{equation} 
in cylindrical polar coordinates. Thus, in the ground state, the external and Hartree potentials due to the underlying lattice are assumed to approximately cancel.  
(In reality, there will be some periodic 
variation in $v_{\mathrm{ext}} + v_{\mathrm{H}}$ in the $z$-direction. However, this contribution will be the same in both the QP and KS descriptions, and, since we are  
interested in the 
differences between the two, such a term can be safely neglected without altering the physics being addressed.)
 
The spin- and current-independent self-energy operator employed \cite{GodbySham1994} has been shown to approximate the $GW$ self-energy of nearly-free electronic 
materials \cite{Godbyetal1988} and is of the form 
\begin{equation} 
 \Sigma(\mathbf{r},\mathbf{r}') = \frac{f(z) + f(z')}{2}g \left( \left| \mathbf{r} - \mathbf{r}' \right| \right) 
\end{equation} 
where $g \left( \left| \mathbf{r} - \mathbf{r}' \right| \right) = \text{exp} \left( - \left( \left| \mathbf{r} - \mathbf{r}' \right| / w \right) ^2 \right)/ \sqrt{\pi} w$  
introduces the nonlocal operation of the self-energy on the quasiparticle wavefunction, while  $f (z) = -F_0 \left[ 1 - \cos(2\pi z/a) \right] $ imposes the  
periodicity along the wire of the underlying crystal lattice. As before, we choose $a = 4$ a.u. and $F_0 = 4.1$ eV. 
The parameter $w$ is chosen as 0.5 a.u. to approximate the Wigner-Seitz radius 
$r_s=(3n/4\pi)^{1/3}$ on the axis of the nanowire. These parameters   
approximately model a one-atom-thick silicon nanowire. For the confining potential, $H = 3$ eV a.u.$^{-6}$  ensures that the charge and current density go to zero  
smoothly at the edge of the wire and that all occupied electron states lie within the first subband.  

We sample the band structure at the $\Gamma$-point for a system of length $L_z = 10a$, thus the quasiparticle is normalized to a supercell of 10 unit cells with each
unit cell contributing one spinless electron to the ground-state charge density, plus one additional electron added to the standing wave state at the bottom of the 
conduction band. The QP is added to the lowest right-going unoccupied eigenstate of the Hamiltonian, yielding a total of 12 electrons per supercell. 

The divergence of the QP current density is given by the continuity equation and the time-dependent form of Eq. \ref{eq:qpequation} 
($E_{\textrm{QP}}\rightarrow i\partial_t$):

\begin{widetext}
\begin{align}
 \nabla \cdot \mathbf{j}(\mathbf{r},t) &= -\frac{\partial}{\partial t}n(\mathbf{r},t) = i\psi^*_{\textrm{QP}}(\mathbf{r},t)\left[ -\tfrac{1}{2}\nabla^2\psi_{\textrm{QP}}(\mathbf{r},t) + \int~d\mathbf{r}'~\Sigma(\mathbf{r},\mathbf{r}')\psi_{\textrm{QP}}(\mathbf{r}',t) \right] + \mbox{c.c} \nonumber \\
 & = \nabla \cdot \mathbf{j}_0(\mathbf{r},t) + 2 \mbox{Re} \int~d\mathbf{r}'~i\psi^*_{\textrm{QP}}(\mathbf{r},t)\Sigma(\mathbf{r},\mathbf{r}')\psi_{\textrm{QP}}(\mathbf{r}',t)
\end{align}
where $\mathbf{j}_0(\mathbf{r},t) = \left< \psi_{\textrm{QP}}(t) \left| \hat{\mathbf{j}}_\textrm{p}(\mathbf{r}) \right| \psi_{\textrm{QP}}(t) \right>$. 
Choosing our axes such that the current density is in the positive $z$-direction gives a steady-state QP current density of
\begin{equation}
\label{eq:qpcurrent}
 \mathbf{j}(\mathbf{r}) = \mathbf{j}_0(\mathbf{r}) - 2\int_{-\infty}^z~dz'~\int~d\mathbf{r}''~\mbox{Im } \psi^*_{\textrm{QP}}(\mathbf{r}')\Sigma(\mathbf{r}',\mathbf{r}'')\psi_{\textrm{QP}}(\mathbf{r}'').
\end{equation}
\end{widetext}
(Note that the paramagnetic current operator acting on the QP wavefunction does {\it not} generally give the paramagnetic current density of the many-body system.) 

One may note that the second term in Eq. \ref{eq:qpcurrent} is zero for the homogeneous electron gas (HEG) for a real-valued and spherically-symmetric self-energy 
operator. As such, the QP current would have no explicit dependence on the nonlocal range of the operator, as one would expect since the QP wavefunction varies spatially 
only in its phase. Thus exact KS-DFT should generally be able to reproduce the current-density of a degenerate HEG. For spatially-varying systems, however, the 
current density will explicitly depend on the $k$-dependent self-energy, and therefore the Coulomb strength of the interacting system, with the gradient of the current 
given by
\begin{align}
 \nabla \cdot \mathbf{j}(\mathbf{r}) &= \nabla \cdot \mathbf{j}_0(\mathbf{r}) \nonumber \\
 &- \frac{2}{V} \mbox{Im } \sum_{\alpha,\beta}c^*_{\alpha}c_{\beta}\sigma(\mathbf{k}_{\beta})\exp{(i(\mathbf{k}_{\alpha}-\mathbf{k}_{\beta})\cdot\mathbf{r})}
\end{align}
where $V$ is the volume of the supercell and
\begin{align}
\psi_{\textrm{QP}}(\mathbf{r}) &= \sum_{\alpha}c_{\alpha}\exp{(i\mathbf{k}_{\alpha}\cdot\mathbf{r})}, \\
\Sigma(\mathbf{r},\mathbf{r}') &= \sum_{\alpha}\sigma(\mathbf{k}_{\alpha})\exp{\left(i \mathbf{k}_{\alpha}\cdot\left( \mathbf{r}-\mathbf{r}'\right)\right)}.
\end{align}

For spatially-varying systems, therefore, the current density depends nonlocally on the system and explicitly on the interaction strength. It is possible, then, that two 
degenerate systems having the same charge density and different self-energy operators will not have the same current density, and therefore not have $V$-representable 
(i.e. representable by a noninteracting {\it scalar} potential only) charge and current densities.

\section{The application of DFT to current-carrying systems}

\begin{figure} 
\includegraphics[scale=0.6]{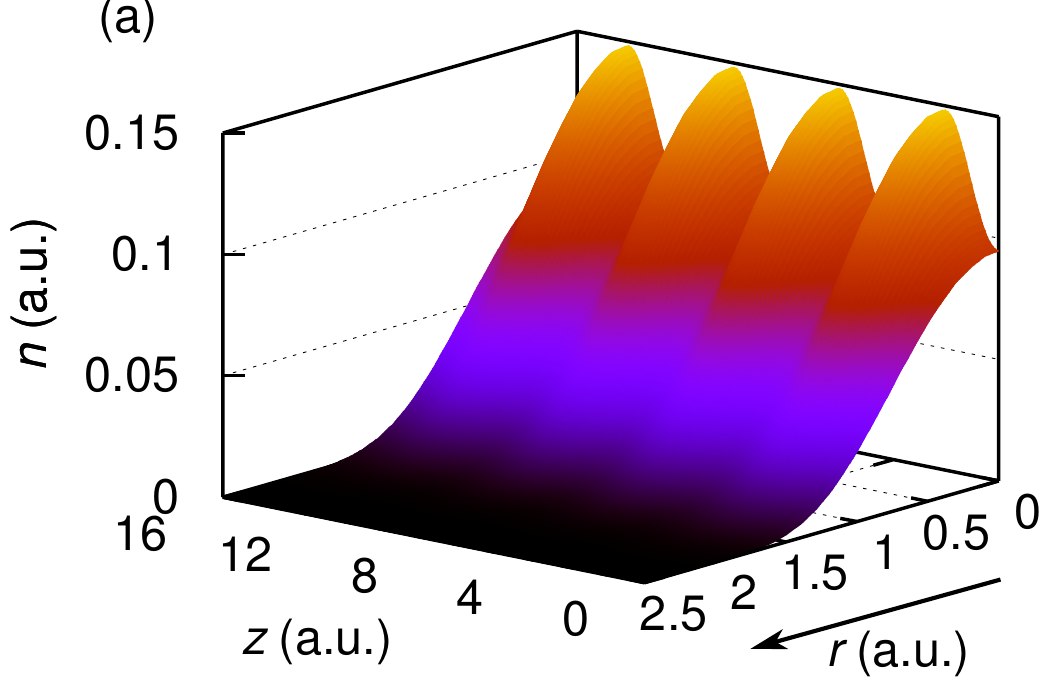} \\ 
\includegraphics[scale=0.6]{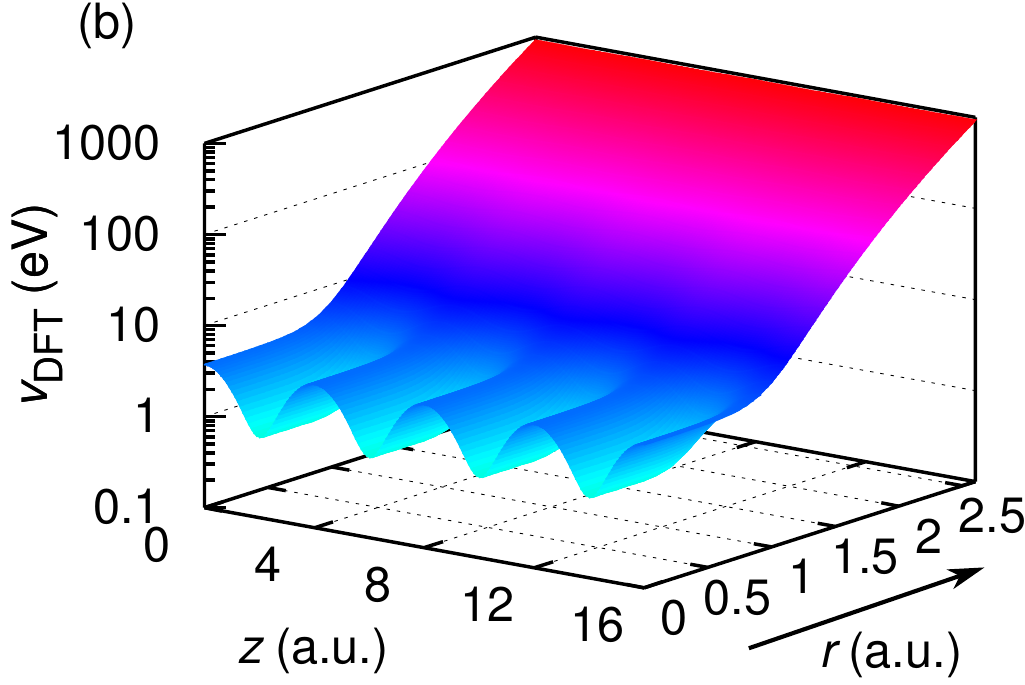} \\ 
\includegraphics[scale=0.53]{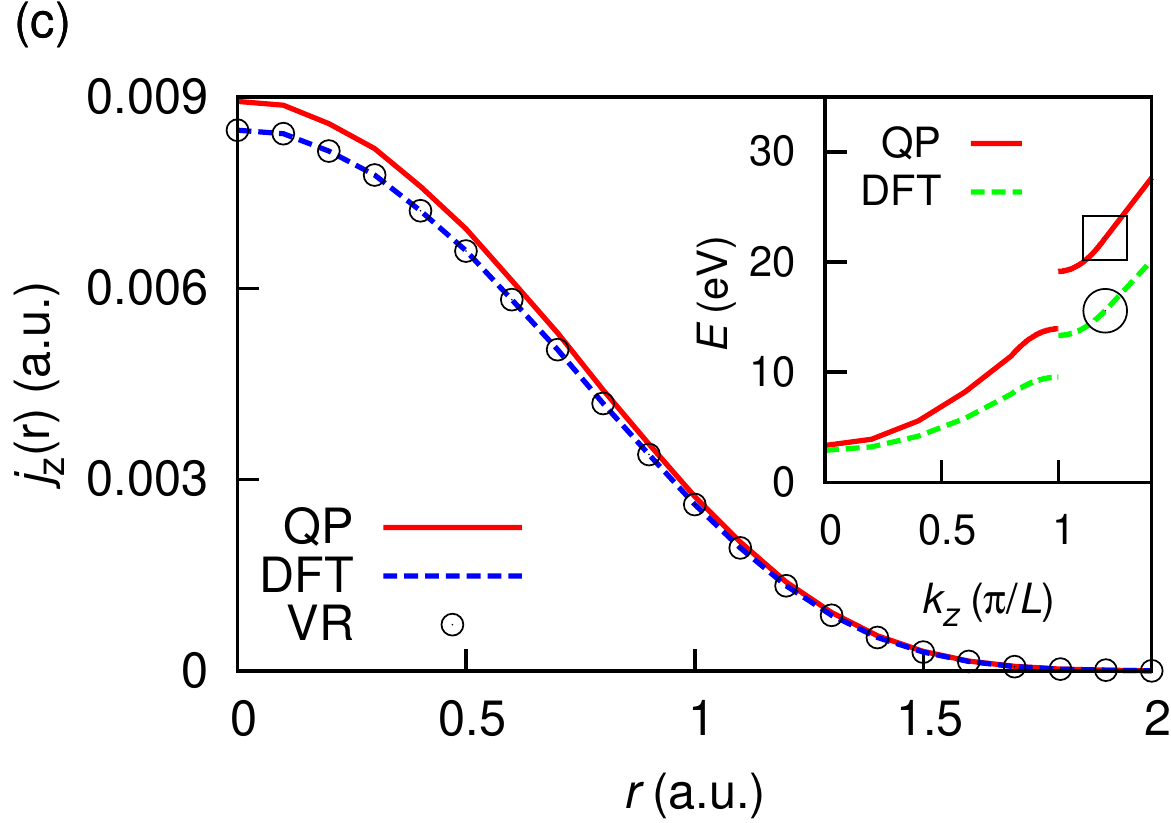} \\
\includegraphics[scale=0.6]{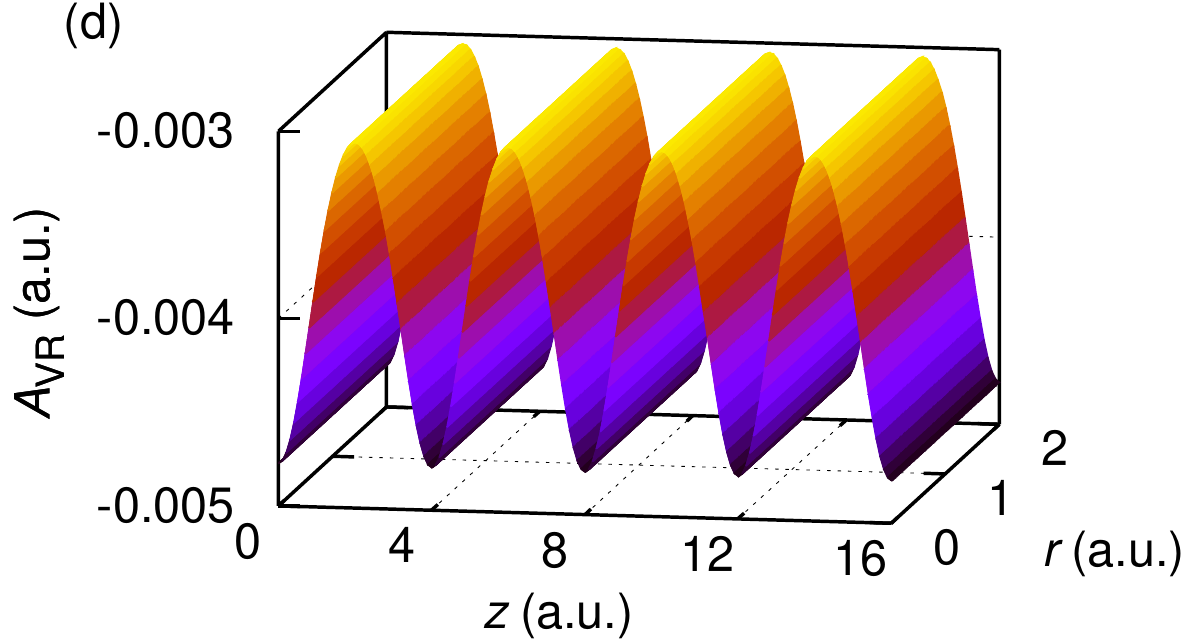}
\caption{\label{fig:rgfig1} (Color online) The axial and radial dependence of (a) the charge density of the system governed by the nonlocal self-energy, and (b) the  
KS scalar potential which reproduces the charge density exactly. The quasiparticle current density (c)  
varies only radially in the steady-state regime. The current density predicted by this DFT scalar potential (blue dashed)  
underestimates the true current density (red solid) by 5\%. The underestimate is due to the difference in band structures (inset) which have different gradients in the  
region of the highest occupied state (denoted by square and circle respectively). In a standard CSDFT approach, the correct paramagnetic current can be obtained
with the inclusion of a KS vector potential (d); however such a potential corresponds only to a gauge transformation: it leaves the charge and physical 
current density (circles) unchanged, and thus gives no improvement over the DFT description of the system.}
\end{figure} 

Figure \ref{fig:rgfig1} shows (a) the charge density of the steady-state system, along with (b) the exact DFT scalar potential which reproduces it in the KS scheme. The 
scalar potential is  calculated first using the van Leeuwen-Baerends \cite{LeeuwenBaerends1994} procedure which iteratively multiplies the potential by 
$n_{\mathrm{KS}}/n$, and is then further  refined by iterative addition of $n_{\mathrm{KS}}-n$, achieving 
an accuracy of $0.005 \% $. The net effect of the additional DFT potential, besides yielding the correct density amplitude along the wire, is to make  
the effective external potential somewhat less confining to reflect the additional electron-electron repulsion. 
 
Also shown is (c) the radial variation of the resultant QP and DFT current densities. Since we are in the steady-state (indeed ground-state) regime, any spatial 
variation in the current density must be due to a purely {\it transverse} component. Without vector potentials, the current densities of both the QP and DFT 
representations are purely paramagnetic. In the steady-state regime, the continuity equation merely constrains
the divergence of the current density to be zero. Since all of the current is in the axial direction, the radial and azimuthal dependence of the current 
must be calculated directly from the wavefunction. The DFT calculation of this current density yields an error of $5.1 \% $ in the 
center of the wire \cite{footnote2} due to the differences in the band structures, and thus the group velocities, of the QP and DFT electrons. 
This reflects the known fact that exact KS-DFT calculations may yield qualitatively inaccurate band structures \cite{GodbyNeeds1989}.

It follows that DFT does not generally yield the correct current density of a ground-state system subject only to an external electric field. 
This is a time-independent variation of the phenomenon observed in time-dependent systems subject only to external scalar potentials demonstrated by D'Agosta and 
Vignale \cite{dAgostaVignale05}. Thus the necessity of moving to 
a current-density functional theory lies not in the presence of an external vector potential, but in the spatial variation of the current-carrying system.

It is worthy of note that there are no magnetic phenomena at all in the model interacting system. The external vector potential is everywhere zero and the 12 electrons 
are spinless. An exact self-energy operator calculated self-consistently from Hedin's equations would not contain current-dependent and the model self-energy employed 
above introduces no such effects.

\section{The paramagnetic current in the KS scheme}

Let us first consider the CSDFT approach of VR, in which KS systems are constructed to yield the same $(n, \mathbf{j}_{\textrm{p}})$ as the interacting systems they 
represent. The Hamiltonian for a KS system of spinless electrons subject to both scalar and vector potentials is
\begin{equation}
 \label{eq:PauliH}
 \hat{H} = \tfrac{1}{2}\left[ \hat{\mathbf{p}} + \mathbf{A}_{\mathrm{KS}}(\mathbf{r}) \right]^2 + v_{\mathrm{KS}}(\mathbf{r}).
\end{equation} 
While the KS wavefunctions are uniquely determined by $(n, \mathbf{j}_{\textrm{p}})$, the potentials $(v_{\textrm{KS}},\mathbf{A}_{\textrm{KS}})$ are not 
\cite{nonuniqueness} and so one cannot generally speak of \textit{the} exact KS vector potential \textit{per se}. Fig. 1(d) shows 
\textit{a} reverse-engineered vector potential that achieves our desired densities (the corresponding scalar potential is simply 
that of the DFT calculation). However, this vector potential corresponds only to a gauge transformation of the KS single-particle wavefunctions of the form
\begin{equation}
 \psi_{\textrm{KS},k}(\mathbf{r}) \leftarrow e^{i\lambda(\mathbf{r})}\psi_{\textrm{KS},k}(\mathbf{r}).
\end{equation}
where $\lambda(\mathbf{r})$ is the scalar field such that $\mathbf{A}(\mathbf{r}) = -\boldsymbol{\nabla}\lambda(\mathbf{r})$. This vector potential ensures that 
$(n, \mathbf{j}_{\textrm{p}})$ of the KS system are those of the interacting system (Fig. 1(c)), but no physical quantity has changed as a result and the physical 
current of the VR KS system is identical to the ordinary DFT prediction. Since the wavefunction is uniquely determined by $(n, \mathbf{j}_{\textrm{p}})$, it follows that 
an alternative choice of $(v_{\textrm{KS}}, \mathbf{A}_{\textrm{KS}})$ that yield the same $(n, \mathbf{j}_{\textrm{p}})$ also corresponds to a gauge-transform of the 
DFT result. Thus, as was the case with DFT, the VR formulation of CSDFT does not in general reproduce the correct physical current.  For instance, it follows that the VR 
theory is not the limit of time-dependent CDFT in the steady-current limit.

\section{The physical current in the KS scheme}

For this reason, we consider instead a Kohn-Sham scheme that reproduces the \textit{physical} current density of the ground-state nanowire. Putting aside the question of 
which basic variables uniquely determine the ground-state properties of a current-carrying system, one may retain from the approaches of Diener and Pan and Sahni the 
convention of constructing the auxiliary KS systems to have the same $(n, \mathbf{j})$ as the interacting system. 
In the absence of an external vector potential, the exact KS vector potential arises purely from exchange and correlation (XC) and can be calculated iteratively using 
\begin{equation} 
 \mathbf{A}_{\mathrm{xc}}(\mathbf{r}) \leftarrow \mathbf{A}_{\mathrm{xc}}(\mathbf{r}) + \frac{\mathbf{j}(\mathbf{r}) - \mathbf{j}_{\mathrm{KS}}(\mathbf{r})}{n(\mathbf{r})}. 
\end{equation} 
Generally, transverse components to the reverse-engineered vector potential will result in a change in the charge density and therefore necessitate a recalculation of the 
scalar potential. However, the stronger the confinement in the transverse direction, the smaller the perturbation of the charge density. As a result, the converged 
CDFT scalar potential is almost exactly that of DFT (Fig. \ref{fig:rgfig1}). We have confirmed that the sensitivity of the charge density to external vector potentials 
returns as one makes the confining electric field weaker.

\begin{figure} 
\includegraphics[scale=0.6]{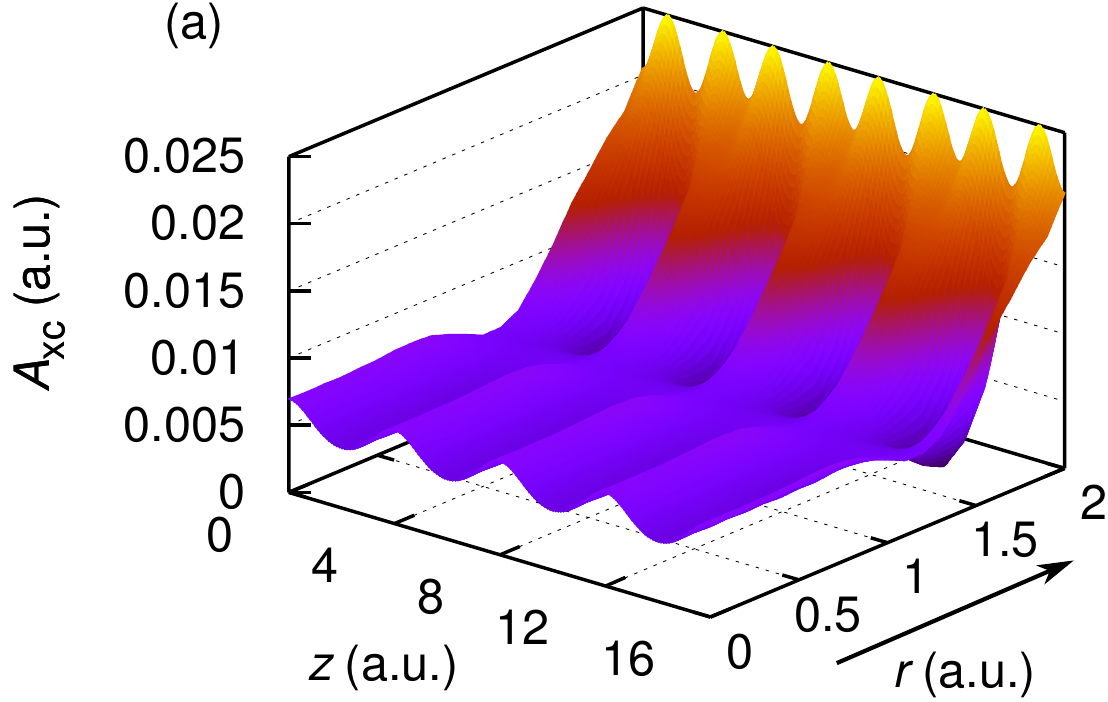} \\ 
\includegraphics[scale=0.6]{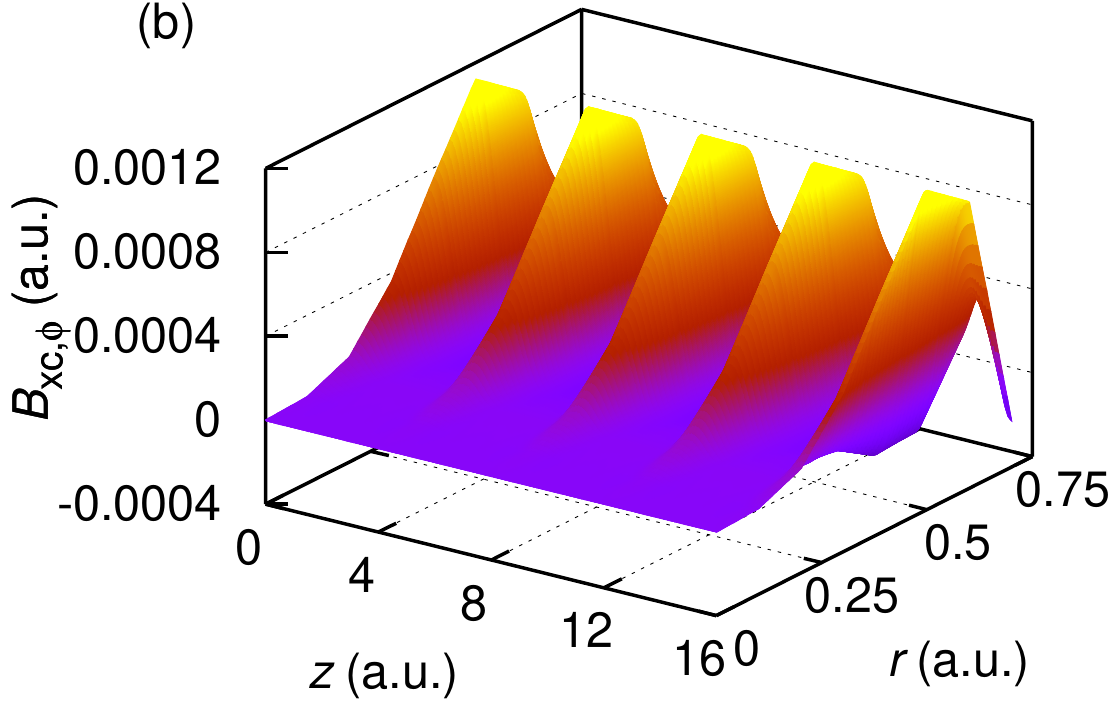} \\ 
\includegraphics[scale=0.6]{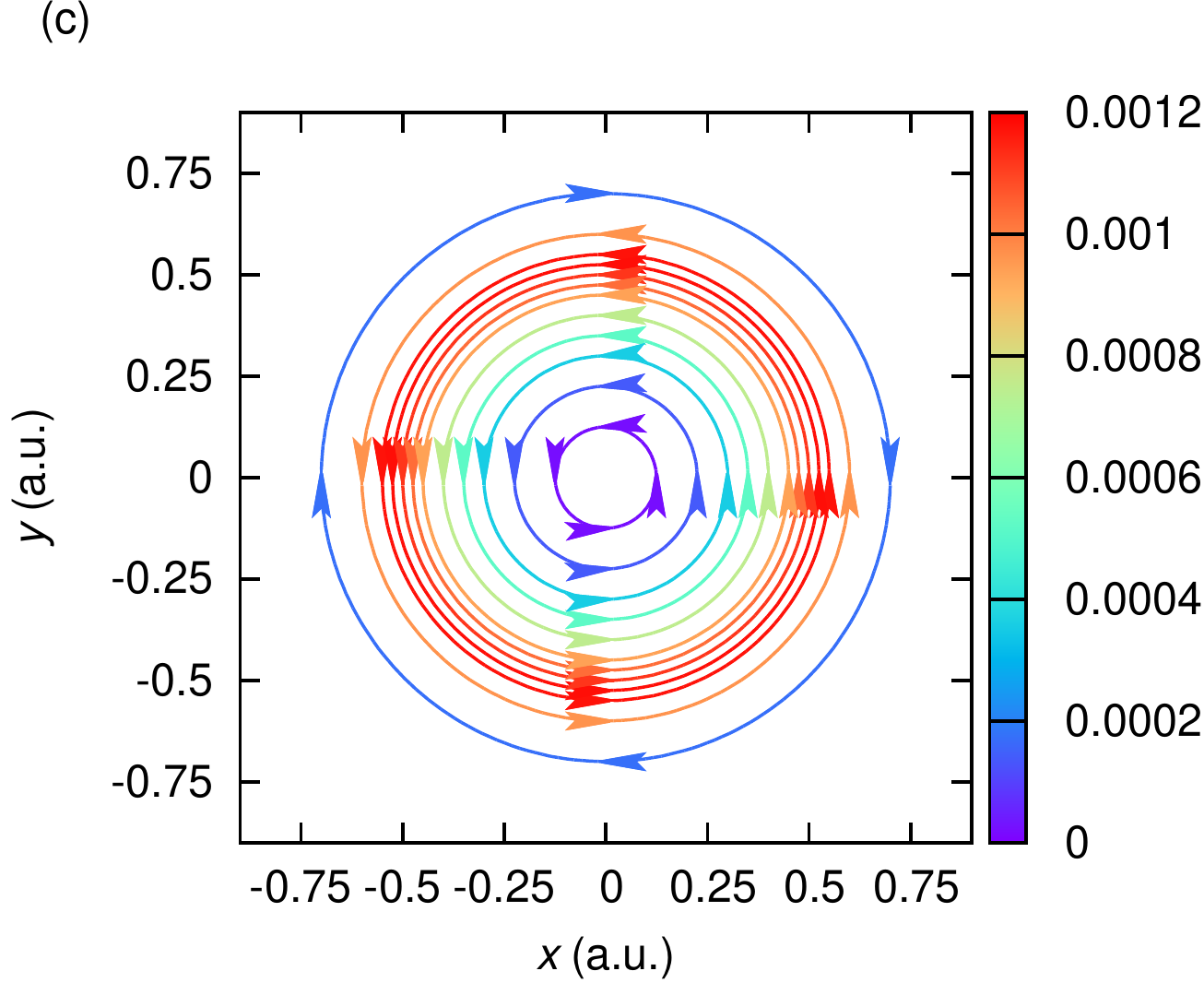} \\ 
\includegraphics[scale=0.6]{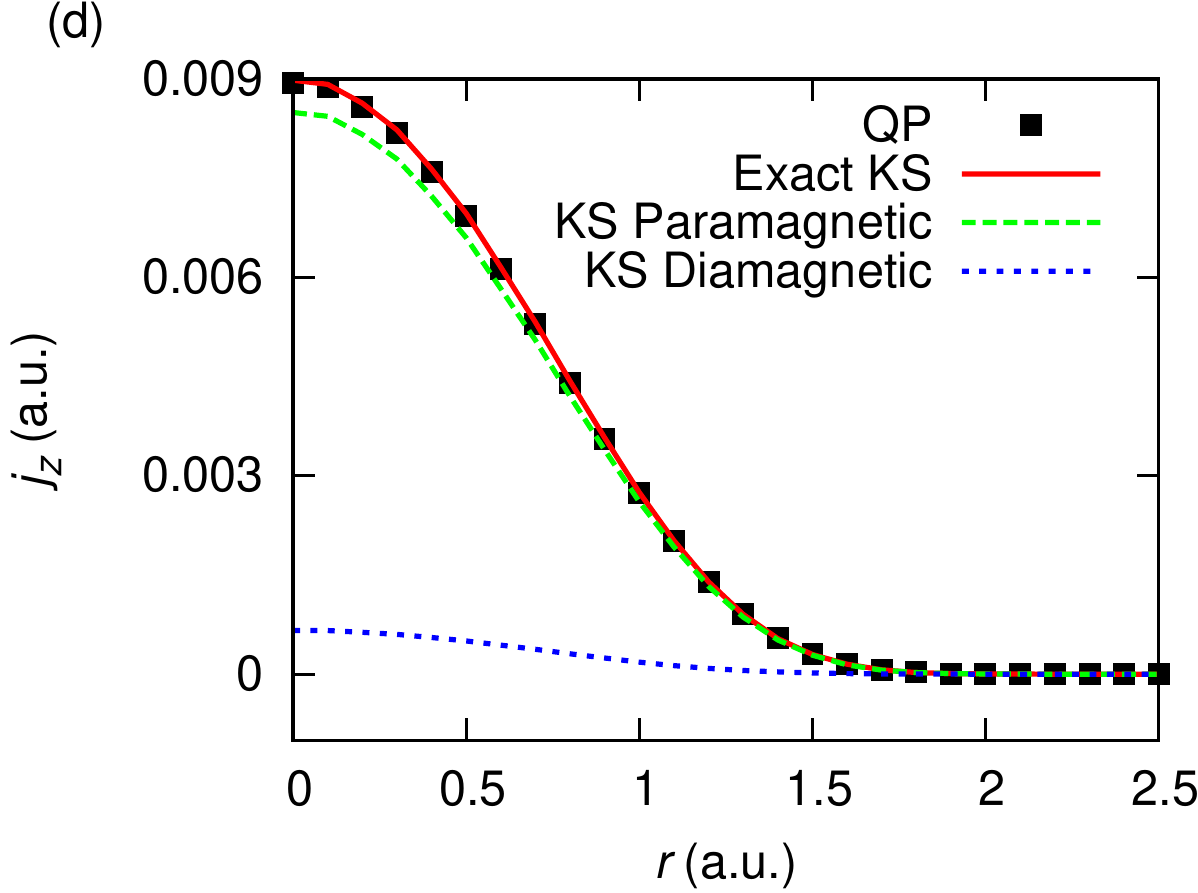} 
\caption{\label{fig:rgfig2} (Color online) (a) The intrinsic KS vector potential and (b) the magnetic field to which it corresponds. These, along with the scalar potential  
of Fig. 1b, reproduce the exact charge and current density of the nanowire. The magnetic field is purely azimuthal and increases with radius (c, shown for $z=0$).  
The resulting KS current density (d, red solid) is now that of the QP system (black squares), but now comprises both paramagnetic (green dashed) and  
diamagnetic (blue dotted) components.} 
\end{figure} 
 
The XC vector potential and the current density that it yields are shown in Fig. \ref{fig:rgfig2}. One can see that, due to the large confining external potential, 
the vector potential becomes large in magnitude at large radii where $n$ and $\mathbf{j}_{\textrm{p}}$ (and thus the effect of the vector potential) is smallest. Because 
of the necessary presence of the nonzero KS vector 
potential, the physical current density in the KS system is no longer purely paramagnetic, in contrast to both the QP and DFT systems. Instead, the physical QP  
current is reproduced by finding the correct combination of paramagnetic and diamagnetic KS current densities for a given charge density. Thus in constructing  
KS functionals, it is necessary to go beyond the paramagnetic current density if one wishes to describe the KS system exactly. Furthermore, because of the  
distribution of the current density between paramagnetic and diamagnetic parts, 12\% of the current density is now carried by the $N$ lowest-energy KS electrons, 
whose counterpart in the QP description of the system contributed zero current.

As in the time-dependent regime, the vector potential calculated corresponds to an exchange-correlation magnetic field 
$\mathbf{B}_{\mathrm{xc}} = \boldsymbol{\nabla} \times \mathbf{A}_{\mathrm{xc}}$, also shown in Fig. \ref{fig:rgfig2}. The manner in which the magnetic field fixes the 
physical current density while keeping the charge density fixed can be characterized as the interplay between two distinct effects.

First, the Kohn-Sham magnetic field is entirely azimuthal, which for an  
axial current corresponds to a \textit{radial}, velocity-dependent force which augments the radial force arising from the Kohn-Sham scalar potential \cite{footnote3}. 
In the current-carrying region, the ratio of the magnetic and electric radial forces (as measured by $u_z B_{\mathrm{xc},\phi} /\partial_r v_{\mathrm{KS}}$)  
is around 2-3\%, comparable to the 5\% adjustment to the current density that CDFT needs to achieve.  The magnitude of the XC  
magnetic field may be gauged from the fact that it is similar in strength to the elementary Biot-Savart magnetic field that the current density  
generates, in the vicinity of the wire.
 
Second, the vector potential ``tunes'' the band structure such that the current density (both paramagnetic and diamagnetic) of all of the current-carrying  
electrons in the KS system, determined in part by the local gradient of the band structure, sum to the correct current of the quasiparticle. 

What is unusual in this case is that the interacting system being modelled does not require the existence of magnetic fields or magnetic interactions at all: the 
electrons are spinless; there are no external magnetic fields applied, and the self-energy operator contains no current- or spin-dependence. The XC vector potential is 
therefore purely mechanical in nature, even though it enters into the Kohn-Sham equations as a magnetic field.

\section{Summary and conclusions}

In conclusion, even in the absence of spin and external magnetic fields, current-carrying systems cannot generally be represented exactly by density functional 
theory and require a purely XC magnetic field that is mechanical in nature and depends on the charge and {\it physical} current density of the system. This XC magnetic 
field arises from the construction of the KS scheme such that the noninteracting physical densities are those of the interacting system, irrespective of the choice of 
basic variables in the underlying CDFT. We have demonstrated a method for the calculation of the necessary exact XC magnetic fields, finding them to be dependent on both 
the charge and current density of the interacting system. Generally, the exact KS representation of a ground-state current-carrying system does \textit{not} carry the 
paramagnetic current density of the interacting system that it represents.

We thank Peter Bokes and Giovanni Vignale for fruitful discussions, and acknowledge funding from EPSRC.


\begin{thebibliography}{99}
\bibitem{footnote2} 
  We have confirmed that, as the nonlocal range, $w$, of the self-energy operator is reduced to zero, the error in the DFT current density also goes to zero.
\bibitem{footnote3}
  In general, the QP will carry a nonzero spin and thus the XC magnetic field will couple to this via the Stern-Gerlach term in Eq. \ref{eq:PauliH}. The Stern-Gerlach 
  interaction will modify the KS charge density from that of the model nanowire without the inclusion of a corresponding modification to the scalar potential. Since 
  the interaction depends on both the XC magnetic field and the spin, the KS scalar potential is necessarily and, in principle, strongly dependent on 
  the current density and the magnetization, as well as the charge density.
\end{thebibliography}
\end{document}